\documentclass[conference]{IEEEtran}
\IEEEoverridecommandlockouts
\usepackage{cite}
\usepackage{amsmath,amssymb,amsfonts}
\usepackage{algorithmic}
\usepackage{graphicx}
\usepackage{textcomp}
\usepackage{xcolor}
\usepackage{import}
\usepackage{svg}
\usepackage{lipsum}
\usepackage{siunitx}
\usepackage{amsmath}
\usepackage{amssymb}
\usepackage{titlesec}

\usepackage{adjustbox}
\usepackage{lengthconvert}
\Convertsetup{unit=cm}
\usepackage{bm}
\usepackage[acronym, shortcuts]{glossaries}

\makeglossaries

\newacronym{DBMC}{DBMC}{diffusion-based molecular communication}
\newacronym{TDMA}{TDMA}{time-division multiple access}
\newacronym{MDMA}{MDMA}{molecule-division multiple access}
\newacronym{NOMA}{NOMA}{non-orthogonal multiple access}
\newacronym{ADMA}{ADMA}{amplitude-division multiple access}
\newacronym{MA}{MA}{multiple access}
\newacronym{BEP}{BEP}{bit error probability}
\newacronym{TX}{TX}{transmitter}
\newacronym{RX}{RX}{receiver}
\newacronym{ISI}{ISI}{inter-symbol interference}
\newacronym{MAI}{MAI}{multiple-access interference}
\newacronym{SIC}{SIC}{successive interference cancellation}
\newacronym{OOK}{OOK}{on-off-keying}
\newacronym{MCS}{MCS}{Monte Carlo simulation}
\newacronym{SNR}{SNR}{signaling-molecule-to-noise ratio}
\newacronym{MC}{MC}{molecular communication}
\newacronym{IoBNT}{IoBNT}{Internet of Bio-Nano-Things}
\newacronym{BNM}{BNM}{bio-nano-machine}
\newacronym{DBMC-NOMA}{DBMC-NOMA}{NOMA for DBMC networks}
\newacronym{UCA}{UCA}{uniform concentration assumption}
\newacronym{CIR}{CIR}{channel impulse response}
\newacronym{SIMO}{SIMO}{single-molecule}
\newacronym{MUMO}{MUMO}{multi-molecule}
\newacronym{bps}{bps}{bits-per-second}
\newacronym{BER}{BER}{bit error rate}

\glsdisablehyper

\def\BibTeX{{\rm B\kern-.05em{\sc i\kern-.025em b}\kern-.08em
    T\kern-.1667em\lower.7ex\hbox{E}\kern-.125emX}}

\usepackage[absolute,showboxes]{textpos}
  
\TPMargin{5pt}
  
\newcommand{\copyrightstatement}{
    \begin{textblock}{0.84}(0.08,0.01)    
         \noindent
         \footnotesize
         \copyright 2024 IEEE. Personal use of this material is permitted. Permission from IEEE must be obtained for all other uses, in any current or future media, including reprinting/republishing this material for advertising or promotional purposes, creating new collective works, for resale or redistribution to servers or lists, or reuse of any copyrighted component of this work in other works.
    \end{textblock}
}
    
\begin{document}

\copyrightstatement

\title{Evaluation of a Multi-Molecule Molecular Communication Testbed Based on Spectral Sensing\
\thanks{The authors acknowledge the financial support by the Federal Ministry of Education and Research of Germany in the program of “Souverän. Digital. Vernetzt.”. Joint project 6G-life, project identification number: 16KISK002.}
}

\author{\IEEEauthorblockN{Alexander Wietfeld, Sebastian Schmidt, Wolfgang Kellerer}
\IEEEauthorblockA{\textit{Chair of Communication Networks} \\
\textit{Technical University of Munich}\\
Munich, Germany\\
\{alexander.wietfeld, sebastian.a.schmidt, wolfgang.kellerer\}@tum.de}
}

\maketitle

\begin{abstract}

This work presents a novel flow-based molecular communication (MC) testbed using spectral sensing and ink intensity estimation to enable real-time multi-molecule (MUMO) transmission. MUMO communication opens up crucial opportunities for increased throughput as well as implementing more complex coding, modulation, and resource allocation strategies for MC testbeds. An estimator using non-invasive spectral sensing at the receiver is proposed based on a simple absorption model. We conduct in-depth channel impulse response (CIR) measurements and a preliminary communication performance evaluation. Additionally, a simple analytical model is used to check the consistency of the CIRs. The results indicate that by utilizing MUMO transmission, on-off-keying, and a simple difference detector, the testbed can achieve up to 3 bits per second for near-error-free communication, which is on par with comparable testbeds that utilize more sophisticated coding or detection methods. Our platform lays the ground for implementing MUMO communication and evaluating various physical layer and networking techniques based on multiple molecule types in future MC testbeds in real time.
\end{abstract}


\begin{figure*}[ht]
    \centering
    \includeinkscape[width=\linewidth]{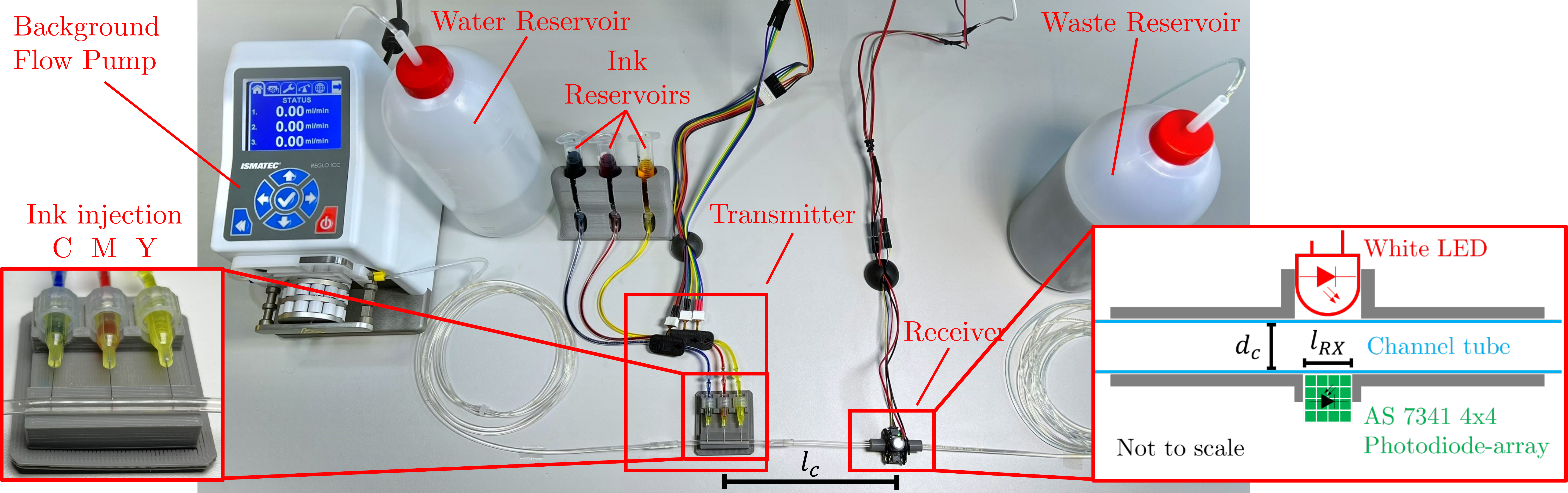}
    \caption{Annotated image of the entire testbed structure. From a water reservoir, the background flow is generated by a peristaltic pump through tubes of diameter $d_\mathrm{c}$. The liquid then flows past the \ac{TX} which consists of three micropumps capable of injecting cyan, magenta, and yellow ink into the channel. After a channel of length $l_\mathrm{c}$ the mixture flows past the non-invasive \ac{RX} with a spectral sensor of length $l_\mathrm{RX}$, and lastly arrives in a waste reservoir.}
    \label{fig:overview}
\end{figure*}

\section{Introduction} \label{sec:introduction}

\Ac{MC} is based on the transfer of molecules for information exchange. It is envisioned to enable various revolutionary use cases in nanotechnology and medicine by providing nano-scale, energy-efficient, and bio-compatible communication, for example, inside the human body.
While research into theoretical models and simulation frameworks has advanced significantly since the emergence of \ac{MC} around two decades ago, there are still only relatively few experimental implementations to validate \ac{MC} systems.

However, some concepts for \ac{MC} testbeds have been proposed in the past~\cite{lotterExperimentalResearchSynthetic2023}.
Among the first approaches for a practical implementation of \ac{MC} were air-based unconstrained setups, where a spray-based \acf{TX} would propel information molecules, often an alcohol solution, towards a receiving sensor~\cite{farsadTabletopMolecularCommunication2013, hofmannTestbedbasedReceiverOptimization2022}. These testbeds are inexpensive, quick to set up, and have successfully been used for more sophisticated communication techniques such as MIMO~\cite{kooMolecularMIMOTheory2016} and network coding~\cite{hofmannAnalogNetworkCoding2023}.
The second large category consists of flow-based experimental setups, often used as abstract representations of future in-body communications, for example, in the bloodstream. It is common to constrain the system to circular tubes with diameters of millimeter scale, induce flow, and transmit information using various types of pumps and information molecules, respectively. 
Farsad \emph{et al.} presented an experimental \ac{MC} platform based on transmitting acid and base as information molecules using multiple peristaltic pumps and inserting a pH probe into the channel as a \acf{RX}~\cite{farsadNovelExperimentalPlatform2017}. The testbed achieves data rates of up to 4 \ac{bps} with a recurrent neural network detector.
In~\cite{brandClosedLoopMolecular2024}, Brand \emph{et al.} implemented a closed-loop flow-based testbed using a switchable fluorescent protein that can enter different states to convey information to the \ac{RX}. The proposed \textit{media modulation} concept increases efficiency since the same set of information molecules can be used for many consecutive transmissions. Two LED arrays for activating and deactivating the protein are used as \ac{TX} and \textit{eraser}, while a fluorescence measurement with a spectrometer occurs at the \ac{RX}. The paper demonstrates a bit rate of 0.1~\ac{bps} over a distance of $\qty{6}{\centi\meter}$ using a difference detector.
In~\cite{angerbauerSalinityBasedMolecularCommunication2023b}, Angerbauer \emph{et al.} describe an implementation of a salinity-based \ac{MC} testbed using a microfluidic chip, micropumps for transmission, and a voltage measurement inside the channel on the \ac{RX} side. Error-free transmission was achieved for bit rates up to 2 \ac{bps} over a channel length of $\qty{3}{\centi\meter}$ using multi-level concentration shift keying and a Viterbi decoder.
The experimental \ac{MC} system built by Wicke \emph{et al.} uses special magnetic nanoparticles as information molecules, a peristaltic pump as a \ac{TX}, and a non-invasive susceptometer as an \ac{RX}~\cite{wickeExperimentalSystemMolecular2022}. In~\cite{bartunikDevelopmentBiocompatibleTestbed2023}, several improvements of the testbed in \cite{wickeExperimentalSystemMolecular2022} are described in detail, including micropumps at the \ac{TX} and a convolutional neural network detector, yielding a bit rate of approximately 6 \ac{bps}. In \cite{wickeExperimentalSystemMolecular2022}, different channel lengths between $\qty{5}{\centi\meter}$ and $\qty{40}{\centi\meter}$ are compared, while the performance bound in \cite{bartunikDevelopmentBiocompatibleTestbed2023} is acquired for a channel length of $\qty{5}{\centi\meter}$.
In~\cite{walterRealtimeSignalProcessing2023}, real-time chemical reactions are utilized to implement an \ac{MC} testbed capable of transmitting over multiple meters 
with a precision-syringe-pump \ac{TX}. Experiments are conducted with symbol periods of around $\qty{30}{\minute}$, but the platform demonstrates working \ac{MC} communication with practical chemical reactions across relatively long distances.


All flow-based testbeds described above focus solely on single-link \ac{MC} between one \ac{TX} and one \ac{RX}. While some platforms are potentially extendable to multiple participants, it is important to choose less invasive options for \ac{TX} and \ac{RX} because, for example, peristaltic pumps in the transmission channel, pH probes, or voltage measurements can add disturbances to the channel, which make adding more nodes difficult. The realization of more complex \ac{MC} use cases relies on the communication between multiple low-capability nodes~\cite{akyildizInternetBioNanoThings2015}. Therefore, the experimental investigation of systems with multiple \acp{TX} or \acp{RX} is crucial. Wang \emph{et al.} have proposed a protocol and novel \ac{MA} scheme for a multi-user \ac{MC} system and conducted an evaluation on an experimental platform based on salinity with up to 4 \acp{TX}~\cite{wangPracticalScalableMolecular2023}. The presented system achieves a per \ac{TX} throughput of up to 1 \ac{bps}. The authors also emulate a parallel \ac{MUMO} transmission but run it as two separate evaluations and do not implement it in real-time, suggesting a practical \ac{MUMO} implementation as important future work. 

The previously described \ac{MC} testbeds are limited to \ac{SIMO} transmission. While in~\cite{walterRealtimeSignalProcessing2023}, multiple molecules are used for the chemical reactions, there is a single channel based on \ac{SIMO} transmission.
Pan \emph{et al.} presented an experimental \ac{MC} platform using an RGB color sensor \ac{RX} and colored ink to transmit information from a single \ac{TX} based on a peristaltic pump and electrical valves \cite{panMolecularCommunicationPlatform2022}. While the authors describe a \ac{MUMO} operation of the testbed as theoretically possible, communication is demonstrated only with one color at a maximum rate of 0.1 \ac{bps}. Calí \emph{et al.} have recently presented a thorough analysis of \textit{graphene quantum dots (GQDs)} as signaling molecules in an \ac{MC} testbed based on a microfluidic chip~\cite{caliExperimentalImplementationMolecule2024a}. GQDs can be synthesized to generate different fluorescence emission spectra, thereby representing different molecule types. The authors describe and evaluate an experimental setup where two types of GQDs are differentiated successfully at a \ac{RX} using principal component analysis of the measured spectral signal. Transmissions take place at a frequency of approximately 1 symbol per minute.

In this paper, we present a novel, inexpensive, and easy-to-setup \ac{MC} testbed capable of differentiating multiple molecules in real time. The latter enables \ac{MUMO} communication, thereby increasing throughput and opening up numerous possibilities for implementing more complex coding, modulation, and resource management schemes. 
We utilize a low-cost micropump \ac{TX} and non-invasive spectral \ac{RX} design, which enables flexible adaptation of the physical network parameters and is easy to extend towards multiple \acp{TX} or \acp{RX} due to its repeatable 3D-printed modular structure. The information molecules in the form of printer ink are also inexpensive and safe to handle.
We propose a linear ink intensity estimator capable of extracting three different simultaneously present colors (cyan, magenta, yellow) from a spectral measurement.
Extensive \ac{CIR} measurements of both \ac{SIMO} and \ac{MUMO} transmissions are presented, and the impact of \ac{MUMO} communication on the channel is discussed. Additionally, we apply a simple analytical model to confirm the consistency of our results.
Lastly, we conduct a performance evaluation of the \ac{MUMO} communication system using a simple difference detector and achieve up to 3 \ac{bps} at a \ac{BER} of $10^{-2}$ over a distance of $>\qty{20}{\centi\meter}$, showing the practical communication potential of the testbed in the context of the related work discussed above.

\section{Testbed Design} \label{sec:testbed}

In the following, we describe the structure of our \ac{MC} testbed by briefly focusing on the channel, \ac{TX} and information molecules, and the \ac{RX}. A photograph of the entire testbed structure is depicted in Figure~\ref{fig:overview}.

The channel consists of a cylindrical tube made from a transparent bio-compatible polymer (Tygon) with a diameter of $d_\mathrm{c} = 2\cdot r_\mathrm{c} = \qty{1.6}{\milli\meter}$.
Background flow $Q_0 = \qty{10}{\milli\liter\per\minute}$ is provided by a peristaltic pump (Reglo ICC) with 12 rollers per channel for minimum pulsatility of the flow speed. It pumps water from the reservoir on the left-hand side through the channel. Then, \ac{TX} and \ac{RX} interact with the channel as described in the following sections, and the resulting wastewater is collected in a waste reservoir on the other side.

\subsection{Information Molecules and Transmitter}
In our testbed, we use colored ink as information molecules. To facilitate \ac{MUMO} transmission using inexpensive materials, we have opted to use cyan (C), magenta (M), and yellow~(Y) printer ink. These colors affect different parts of the visible light spectrum, and therefore, their impact is differentiable at the \ac{RX}, as we will show later. We note, that printer inks are used as placeholders for other types of molecules that are tailored towards specific applications, as the inks are of course not themselves suitable, for example, for in-body usage. While colored particles like inks could be applicable in out-of-body use cases like labs-on-chip~\cite{bartunikColourspecificMicrofluidicDroplet2020} without stringent biocompatibility constraints, magnetic nanoparticles have been proposed and utilized as a biocompatible particle type to realize various medical use cases~\cite{wickeMagneticNanoparticleBasedMolecular2019}.

The \ac{TX} consists of three micropumps, which use piezoelectric components to inject small amounts of liquid into the channel. Each pump is connected to an ink reservoir contained in a 3D-printed housing on one end. On the other end, the \ac{TX} interfaces with the channel through a blunt-end syringe needle for each color with an outer and inner diameter of $\qty{0.23}{\milli\meter}$ and $\qty{0.11}{\milli\meter}$, respectively. Another 3D-printed housing ensures consistent alignment of the needles with the channels such that they are inserted perpendicular to the background flow direction. The distance between each needle along the channel is $\qty{1.3}{\centi\meter}$, and the \ac{TX} midpoint is $l_\mathrm{c} = \qty{23.5}{\centi\meter}$ from the \ac{RX} midpoint. The needle injecting cyan is furthest upstream, magenta in the middle, and yellow the furthest downstream. To minimize the influence of downstream needles on the ink injected by the upstream needles, it is crucial that their diameter is very small and that the insertion depth of the needles is decreased along the flow direction. For example, the cyan needle is inserted furthest so that the cyan ink passes by the other needles.
One \ac{TX}, consisting of three pumps, is connected to a single driver chip connected to an Arduino Nano. The pumps are controlled through an input voltage level that determines the flow speed at the outlet. The peristaltic pump causes significant pressure to be applied to the syringe needle outlets. Backflow in the form of water enters the \ac{TX} if no voltage is applied to the micropumps while the background flow is running, a phenomenon observed in \ac{MC} testbeds before~\cite{bartunikDevelopmentBiocompatibleTestbed2023}. Therefore, we define an idle voltage level $V_\mathrm{idle}$, which is always applied to the \ac{TX} micropumps. 

\subsection{Receiver}
The heart of the \ac{RX} is an Adafruit AS7341 spectral sensor. It consists of a 4-by-4 sensor array that measures eight overlapping channels with center wavelengths $\qty{415}{\nano\meter}$, $\qty{445}{\nano\meter}$, $\qty{480}{\nano\meter}$, $\qty{515}{\nano\meter}$, $\qty{555}{\nano\meter}$, $\qty{590}{\nano\meter}$, $\qty{630}{\nano\meter}$, and $\qty{680}{\nano\meter}$. To maximize sensitivity and minimize outside interference, we have created a custom 3D-printed housing, as depicted and annotated on the right-hand side of Figure~\ref{fig:overview}, into which the channel tube can be inserted. A white LED illuminates the channel from the side opposite the sensor. The sensor and LED are connected and controlled by an Arduino Nano, which samples the sensor readings at a frequency of approximately $\qty{19}{\hertz}$.

\section{Ink Intensity Estimation} \label{sec:concentration}
Many detection algorithms in \acrshort{MC} rely on molecule concentrations as input signals. Since the spectral sensor does not directly provide the cyan, magenta, and yellow ink concentrations, $c_\mathrm{c}(t)$, $c_\mathrm{m}(t)$, and $c_\mathrm{y}(t)$, the testbed relies on estimating the individual color intensities from the mixed spectrum measured at the \ac{RX} relative to a calibration measurement. 
In our setup, deviations in the received spectrum are caused by light absorption of the inks passing through the \ac{RX}. The Beer-Lambert law describes the linear relationship between color intensity $c_i(t)$ ($i \in \{ \mathrm{c}, \mathrm{m}, \mathrm{y} \}$) and absorbance $a_i^j(t)$ due to ink $i$ in the $j$-th spectral channel of the \ac{RX} as
\begin{equation} \label{eq:absorbance_single}
    a_i^j(t) = \log_{10} \left( \frac{I_0^j}{I^j(t)} \right) = b_i^j c_i(t),
\end{equation}
where $b_i^j$ is the absorption coefficient of ink $i$ in channel $j$, $I^j(t)$ is the time-varying light intensity measured in the $j$-th channel, and $I_0^j$ is a reference measurement with no ink in the channel~\cite{swinehartBeerLambertLaw1962}\footnote{The reference measurement is obtained by averaging the light intensity in the absence of ink in each channel over several seconds.}. In the presence of multiple inks, the measured absorbance in channel $j$, $a^j(t)$, is the sum of the individual absorbances $a_i^j(t)$~\cite{QuantitiveAnalysis2007} and therefore, in our case:
\begin{equation} \label{eq:absorbance_sum}
    a^j(t) = a_\mathrm{c}^j(t) + a_\mathrm{m}^j(t) + a_\mathrm{y}^j(t).
\end{equation}
Since we measure 8 distinct channels at the \ac{RX}, as described in Section~\ref{sec:testbed}, we can define the absorbance vector 
\begin{equation} \label{eq:absorbance_vector}
    \mathbf{a}(t) = \left[ a^1(t), a^2(t), \cdots, a^8(t)\right]^T,
\end{equation}
the absorption coefficient vector $\mathbf{b}_i$ for ink $i$, and an absorption matrix $\mathbf{B}$
\begin{equation} \label{eq:absorption}
    \mathbf{b}_i = \left[ b_i^1, b_i^2, \cdots, b_i^8\right]^T \text{and}\ \mathbf{B} = \left[\mathbf{b}_\mathrm{c}, \mathbf{b}_\mathrm{m}, \mathbf{b}_\mathrm{y}\right].
\end{equation}
Using Eqs.~(\ref{eq:absorbance_single}) and~(\ref{eq:absorbance_sum}), the previous definitions, and the vector $\mathbf{c}(t) = \left[ c_\mathrm{c}(t), c_\mathrm{m}(t), c_\mathrm{y}(t) \right]^T$, we arrive at the following system of linear equations:
\begin{equation}
\label{eq:system}
    \mathbf{a}(t) = \mathbf{B}\cdot \mathbf{c}(t).
\end{equation}
Due to the linearity of the system, by sending $N_\mathrm{p}$ calibration pulses with predefined peak intensity $c_{i,\mathrm{peak}} = 1$ from each ink individually and averaging the peak absorption, we can determine the $\mathbf{b}_i$ as 
\begin{equation} \label{eq:avg_vector}
    \mathbf{b}_i = \frac{1}{N_{p}} \sum_{k = 1}^{N_{p}}\ \underset{t \in T_k}{\max} \left( \mathbf{a}(t) \right),
\end{equation}
where $T_k$ is a time interval, in which the $k$-th pulse is received. 

To calculate the estimated ink intensity vector $\hat{\mathbf{c}}(t)$ from an absorbance measurement $\mathbf{a}(t)$, we employ ordinary least squares estimation. According to~\cite[Chapter~4]{kayFundamentalsStatisticalSignal1993}, for the overdetermined system in Eq.~(\ref{eq:system}), the least squares optimal estimator is equal to the Moore-Penrose pseudoinverse $\mathbf{B}^\dagger$ and therefore
\begin{equation}
    \hat{\mathbf{c}}(t) = \mathbf{B}^\dagger \mathbf{a}(t) = \left(\mathbf{B}^T\mathbf{B} \right)^{-1}\mathbf{B}^T \mathbf{a}(t).
\end{equation}

\section{Channel Impulse Response Modeling} \label{sec:modeling}
The propagation of the ink information molecules in our testbed is primarily based on advection via bulk background flow of water. Therefore, we will briefly analyze the expected flow properties in our testbed given its parameters and introduce a simple analytical model.

Firstly, we try to categorize the fluid flow in the system as either laminar or turbulent, as this significantly impacts the subsequent modeling~\cite{jamaliChannelModelingDiffusive2019}. The Reynolds number $\mathrm{Re}$ is a dimensionless quantity used to differentiate the laminar and turbulent case and is defined as~\cite[Chapter~1]{whiteFluidMechanics2016}
\begin{equation}
    \mathrm{Re} = \frac{d_\mathrm{c}v_\mathrm{avg}}{\nu},
\end{equation}
with the channel diameter $d_\mathrm{c}$ and the average background flow velocity $v_\mathrm{avg} = \frac{Q_\mathrm{0}}{\pi r_\mathrm{c}^2}$ as the characteristic length and velocity scales of the system, respectively, and the kinematic viscosity of the fluid $\nu$. The transition from laminar to turbulent flow in a smooth circular pipe occurs at approximately $\mathrm{Re} = 2100$~\cite[Chapter~4]{whiteFluidMechanics2016}. Setting $d_\mathrm{c}= \qty{1.6}{\milli\meter}$, $Q_\mathrm{0} = \qty{10}{\milli\liter\per\minute}$, and using the kinematic viscosity of water $\nu_\mathrm{water} = \qty{1.01e-6}{\meter\squared\per\second}$, we get $\mathrm{Re} = 131 \ll 2100$ for our testbed, which we therefore assume to operate in the laminar regime.
For laminar flow in a straight, circular pipe, we can assume a \textit{Poiseuille} flow velocity profile~\cite[Chapter~4]{whiteFluidMechanics2016}. The flow velocity $v$ is a function of the radius $\rho$, radially symmetric around the center of the channel ($\rho=0$), and defined as follows:
\begin{equation}
    v(\rho) = v_\mathrm{max} \left(1-\frac{\rho^2}{r_\mathrm{c}^2}\right),\;\; \rho\in \left[0,r_\mathrm{c}\right],
\end{equation}
where $v_\mathrm{max} = 2\cdot v_\mathrm{avg}$ is the maximum flow velocity in the middle and $r_\mathrm{c} = \frac{d_\mathrm{c}}{2} = \qty{0.8}{\milli\meter}$ is the channel radius.

Secondly, we determine the relative influence of diffusion on the fluid transport, which can be described by the dimensionless Péclet number $\mathrm{Pe} = \frac{r_\mathrm{c} v_\mathrm{avg}}{D}$~\cite{jamaliChannelModelingDiffusive2019}, with the diffusion coefficient $D$. We use dye-based fully dissolved ink and therefore use the diffusion coefficient of water in water, $D \lesssim \qty{2.299e-9}{\meter\squared\per\second}$~\cite{holzTemperaturedependentSelfdiffusionCoefficients2000}, as an upper limit approximation. We can, therefore, calculate $\mathrm{Pe} \gtrsim \qty{2.88e4}{} \gg 1$ and conclude that flow dominates over diffusion, which can be neglected as a result.

For the purposes of a simplified analytical model, we assume the \ac{RX} to be passively observing the molecules over the channel cross-section along a section with length $l_\mathrm{RX}$ centered at a distance $l_\mathrm{c}$ from the \acp{TX}. Additionally, we assume that the \acp{TX} emit an instantaneous pulse of molecules distributed over the cross-section at the injection point. Using the model proposed by Wicke \emph{et al.} in~\cite{wickeExperimentalSystemMolecular2022}, we define a radially symmetric initial distribution based on the Beta distribution
\begin{equation} \label{eq:init_distr}
    f(s) = \frac{\Gamma(\alpha\!+\!\beta)}{\Gamma(\alpha)\Gamma(\beta)} s^{\alpha-1} (1-s)^{\beta-1};\ s = \frac{\rho^2}{r_\mathrm{c}^2} \in [0,1],
\end{equation}
with the Gamma function $\Gamma(\cdot)$ and two variable parameters $\alpha$ and $\beta$. We then assume that $l_\mathrm{c}\gg l_\mathrm{RX}$\footnote{This is valid since $l_\mathrm{c} = \qty{23.5}{\centi\meter}$ and the size of the sensor array inlet is $l_\mathrm{RX} = \qty{0.9}{\milli\meter}$. The effective $l_\mathrm{RX,eff}$ of the observed section will be larger, due to effects such as scattering, but still much smaller than $l_\mathrm{c}$.}, and therefore obtain the \ac{CIR}~\cite{wickeExperimentalSystemMolecular2022}
\begin{equation} \label{eq:analytical_cir}
    h(t)\! =\! \begin{cases}
        C \frac{\Gamma(\alpha+\beta)}{\Gamma(\alpha)\Gamma(\beta)}\!\left(\!1\!-\!\frac{l_\mathrm{c}}{v_\mathrm{max}t}\!\right)^{\alpha-1}\!\left(\!\frac{l_\mathrm{c}}{v_\mathrm{max}t}\!\right)^\beta\!\! & t \geq \frac{l_\mathrm{c}}{v_\mathrm{max}}\\
        0 & \mathrm{else}, 
    \end{cases}
\end{equation}
with a dimensionless scaling factor $C$. Mirroring~\cite{wickeExperimentalSystemMolecular2022}, Eqs.~(\ref{eq:init_distr}) and~(\ref{eq:analytical_cir}) will be used to fit $\alpha$ and $\beta$ towards the measured \ac{CIR} and thereby check the consistency of the \ac{CIR} and initial distribution across trials and between different colors.

\section{Results} \label{sec:results}
In the following section, we will present evaluation results based on measurements from our testbed as described in Section~\ref{sec:testbed} and the methods and models introduced in Sections~\ref{sec:concentration} and~\ref{sec:modeling}. This includes a characterization of the \ac{CIR} and an analysis of the communication capabilities using a random bit sequence. The parameters used throughout the evaluation can be seen in Table~\ref{tab:parameters}.

\begin{table}[]
\caption{Evaluation Parameters}
\label{tab:parameters}
\centering
\resizebox{0.8\columnwidth}{!}{%
\begin{tabular}{lll}
\hline
\textbf{Parameter} & \textbf{Symbol}     & \textbf{Value(s)}                   \\ \hline
Idle pump voltage       & $V_\mathrm{idle}$   & $\qty{40}{\volt}$                   \\
Bit-1 pump voltage      & $V_1$               & $\qty{110}{\volt}$                  \\
Bit-0 pump voltage      & $V_0$               & $\qty{40}{\volt}$                   \\
Symbol period      & $T_\mathrm{sym}$ & $\{0.5, 1, 2, 10\}\ \qty{}{\second}$    \\
Injection time     & $T_\mathrm{inj}$    & $\{0.05, 0.1\}\ \qty{}{\second}$    \\
Background flow    & $Q_0$               & $\qty{10}{\milli\liter\per\minute}$ \\
Channel diameter   & $d_\mathrm{c}$      & $\qty{1.6}{\milli\meter}$ \\
Channel length     & $l_\mathrm{c}$      & $\qty{23.5}{\centi\meter}$ \\ \hline
\end{tabular}%
}
\end{table}

\subsection{Channel Impulse Response Evaluation}
We will use experimental data from the testbed to generate average \ac{CIR} curves for \ac{SIMO} and \ac{MUMO} transmission, respectively. Therefore, we have gathered measurements of 210 pulses in total, 30 trials each of the seven possible molecule combinations from 3 colors, i.e. cyan (C), magenta (M), and yellow (Y), individually, and CM, CY, MY, and CMY. The pulses were generated with an injection time of $T_\mathrm{inj} = \qty{0.1}{\second}$ and separated by $T_\mathrm{sym} = \qty{10}{\second}$ to avoid \ac{ISI}. The peak spectral deviations from the \ac{SIMO} trials were then used to calculate the averaged absorption vectors $\mathbf{b}_c$, $\mathbf{b}_m$, and $\mathbf{b}_y$, as shown in Eq.~(\ref{eq:avg_vector}) with $N_\mathrm{p} = 30$. Subsequently, the absorption matrix~$\mathbf{B}$ is calculated for the intensity estimator as defined in Eq.~(\ref{eq:absorption}). The estimator converts the spectral measurements from each pulse into estimated color intensities. 

\begin{figure}[t]
    \includeinkscape[width=0.325\linewidth]{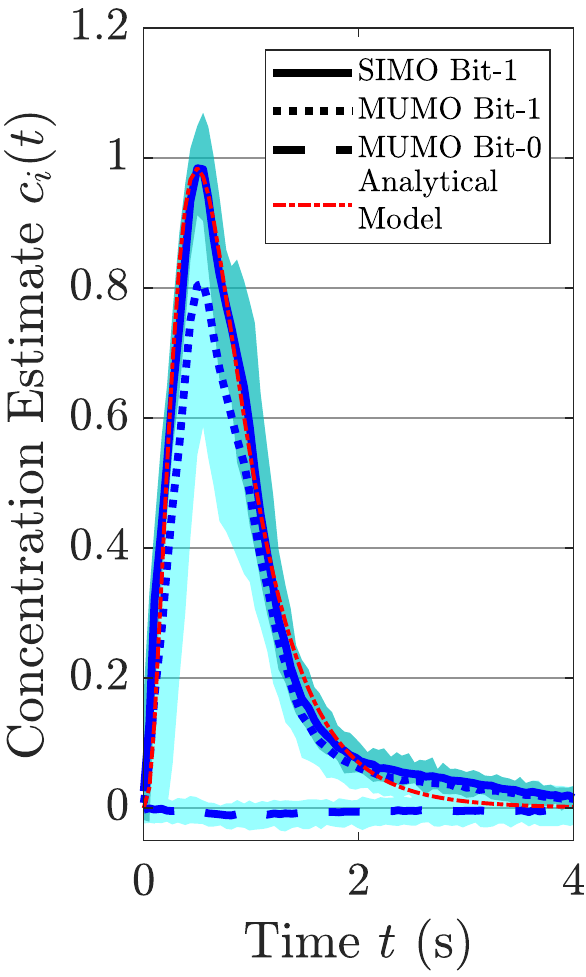} \hfill
    \includeinkscape[width=0.325\linewidth]{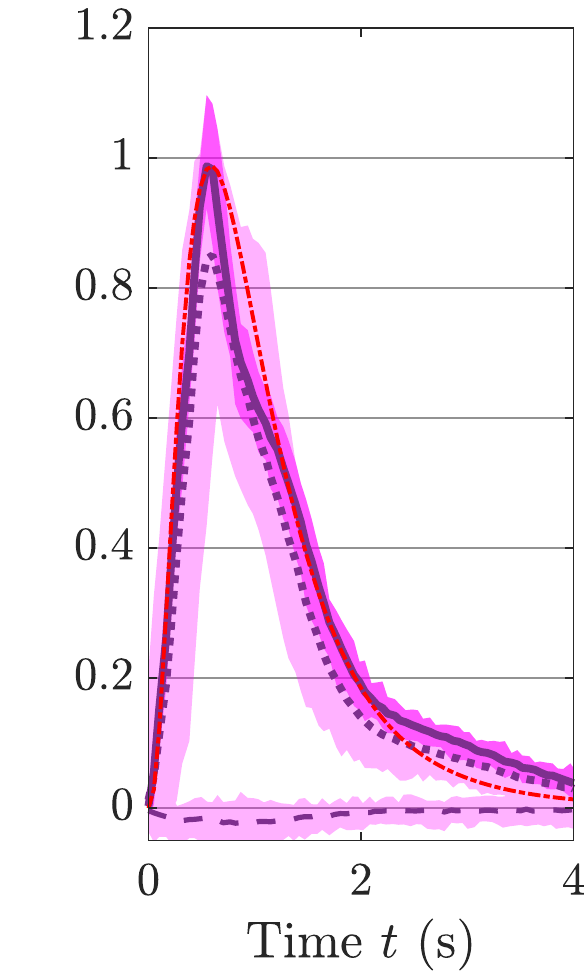} \hfill
    \includeinkscape[width=0.325\linewidth]{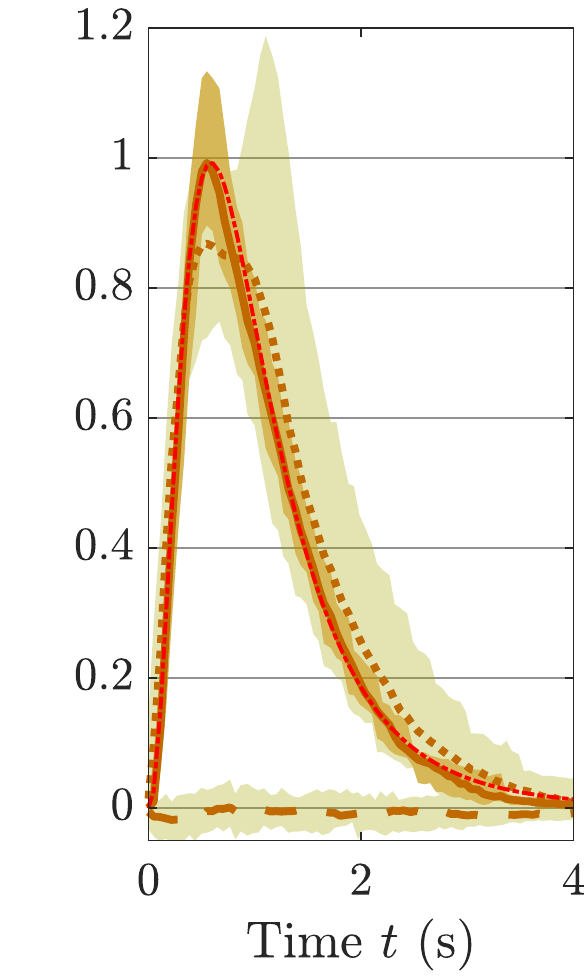}
    \caption{Average \ac{CIR} for \ac{SIMO} and \ac{MUMO} transmission of cyan, magenta, and yellow molecules as the estimated color intensity over time. The plot shows the \ac{SIMO} \ac{CIR} average, with the dark-shaded area representing the maximum deviation from the average. The best fit analytical \ac{CIR} defined in Eq.~(\ref{eq:analytical_cir}) is shown in red. Additionally, the average \ac{MUMO} transmission bit-1 and bit-0 response are shown with the lighter shaded area depicting the maximum deviation from the average.}
    \label{fig:avg_cir}
\end{figure}

Figure~\ref{fig:avg_cir} depicts the results of the averaged \ac{CIR} analysis. The solid C, M, and Y line depicts the averaged \ac{CIR} from the respective \ac{SIMO} pulses, and the darker shaded area around the curve represents the maximum deviation from the average for the \ac{SIMO} pulses. We can observe that they exhibit very similar characteristics in terms of their peak and fall-off time between all three colors, except for minor shape variations. Additionally, the deviations are quite close to the average. All considered \acp{CIR} return to zero after approximately 5 seconds. 

\begin{figure}[t]
    \centering
    \includeinkscape[width=0.8\linewidth]{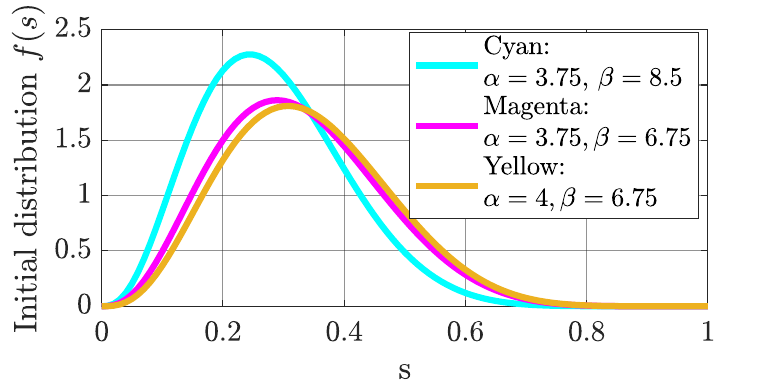}
    \caption{Best fit initial Beta distributions $f(s)$ as defined in Eq.~\ref{eq:init_distr} for the \ac{SIMO} \acp{CIR} shown in Figure~\ref{fig:avg_cir}.}
    \label{fig:init_distr}
\end{figure}

The thin dash-dotted red line in each plot corresponds to the analytical model introduced in Section~\ref{sec:modeling} fitted towards the shape of the \ac{SIMO} \ac{CIR} of each color using exhaustive search and finding the minimum mean-squared error solution for $\alpha$ and $\beta$. We can observe a good fit of the \ac{CIR} shape for the optimal solution in each case, capturing the characteristics of the rise and peak of the \ac{CIR} well while mostly deviating in the tail section. This could be explained by the inaccurate assumption of instantaneous release of the molecules, as small amounts of ink continue to leak out of the injection needle after $T_\mathrm{inj}$ has passed. This would cause an underestimation of the analytical model in the latter part of the \ac{CIR} due to molecules arriving later than expected. The resulting initial distributions and the values for the optimal $\alpha$ and $\beta$ are shown in Figure~\ref{fig:init_distr}. Here, we can see that the fitted initial distributions are similar for all three colors, with M and Y exhibiting near identical results, while C deviates more and, as a result, exhibits a slightly sharper \ac{CIR} peak. The differences here are presumably due to actual deviations in the placement of the injection needles in the channel as described in Section~\ref{sec:testbed} and to imperfections in the physical injection process, such as leakage after the pulse and disturbances of the pulse from other needles in the channel.

Figure~\ref{fig:avg_cir} also depicts the averaged \ac{CIR} for the \ac{MUMO} transmission as a dotted line, encompassing all measured pulses from one color as they are transmitted in parallel to 0, 1, or 2 other colors. The lighter-shaded area around the curve represents the maximum deviation from that average in either direction. In general, the average shape of the curve is retained in the \ac{MUMO} case for all colors. However, we can observe some changes. Firstly, the variations around the average increase significantly, and secondly, the average curve is decreased from the \ac{SIMO} average by about 20\% at the peak across all three colors. There are several contributing factors to this effect. Since the colors are not perfectly mixed when they arrive at the \ac{RX}, the random layering and occlusion of some colors over others could impact how strongly the color is seen on the spectrum due to the directional illumination setup of the \ac{RX} from top to bottom. This would cause a decrease in the measured intensity for some pulses. Additionally, the synchronization and alignment of the pulses across different batches of measurements for each color combination are approximate. The differing distances of the injection needles also cause varying travel times of the colors from the \ac{TX}. Therefore, the \ac{MUMO} averaged pulses include some inherent misalignment, evident in the yellow \ac{CIR} but present for all colors, which broadens and lowers the peak. Overall, we believe the physical setup and its inaccuracies are the leading cause of the differences between \ac{SIMO} and \ac{MUMO} \ac{CIR}.

Figure~\ref{fig:avg_cir} also includes the \ac{MUMO} bit-0 as a dashed line, i.e., the measured response in each color when any other colors are currently being received. Here, the intensity estimate stays close to zero in all cases with no cross-talk from other colors. In general, the \ac{MUMO} \acp{CIR} for bit-1 and bit-0 exhibit a level of consistency high enough for reliable communication, as we will show in Section~\ref{subsec:communication}.

\subsection{Communication System Performance} \label{subsec:communication}

A random balanced sequence of 3 times 100 bits, one 100-bit sequence for each color, was generated for conducting a preliminary examination of the communication capabilities of the testbed. The \ac{TX} uses \ac{OOK} to transmit the bits such that the bit from ink $i$ in symbol period $k$ is $s_{i,k} \in \{0,1\}$. In each symbol period, the bits encoded in the three inks are transmitted simultaneously. At the \ac{RX}, we implemented a simple difference detector to enable operation under the influence of \ac{ISI}. To determine the estimated symbol $\hat{s}_{i,k}$, the detector identifies two samples: the sample acquired at the first sample time $t_0^k$ within symbol period $k$, $c_i(t_0^k)$, and the sample with the highest value within the symbol $\underset{t \in T_k}{\max}\ c_i(t)$. If we define $\Delta_{i,k} = \underset{t \in T_k}{\max}\ c_i(t) - c_i(t_0^k)$, the decision rule is
\begin{equation} \label{eq:slope_detection}
       \hat{s}_{i,k} = \begin{cases}
    1 & \Delta_{i,k} \geq \tau \\
    0 & \Delta_{i,k} < \tau \end{cases}. 
\end{equation}
The detection process is applied simultaneously for all three colors using the same $\tau$ value. A preamble of 4 bit-1s is sent before the sequence to make it easier to identify the start of the transmission as well as the symbol period via the successive peaks. By identifying the beginning of the transmission and the frequency of the pulses, the \ac{RX} can be synchronized with the \acp{TX} before the main bit sequence is sent.
The system \ac{BER} within the 312 total transmitted bits across all colors is used for the performance evaluation.

\begin{figure}[t]
    \centering
    \includeinkscape[width=0.75\linewidth]{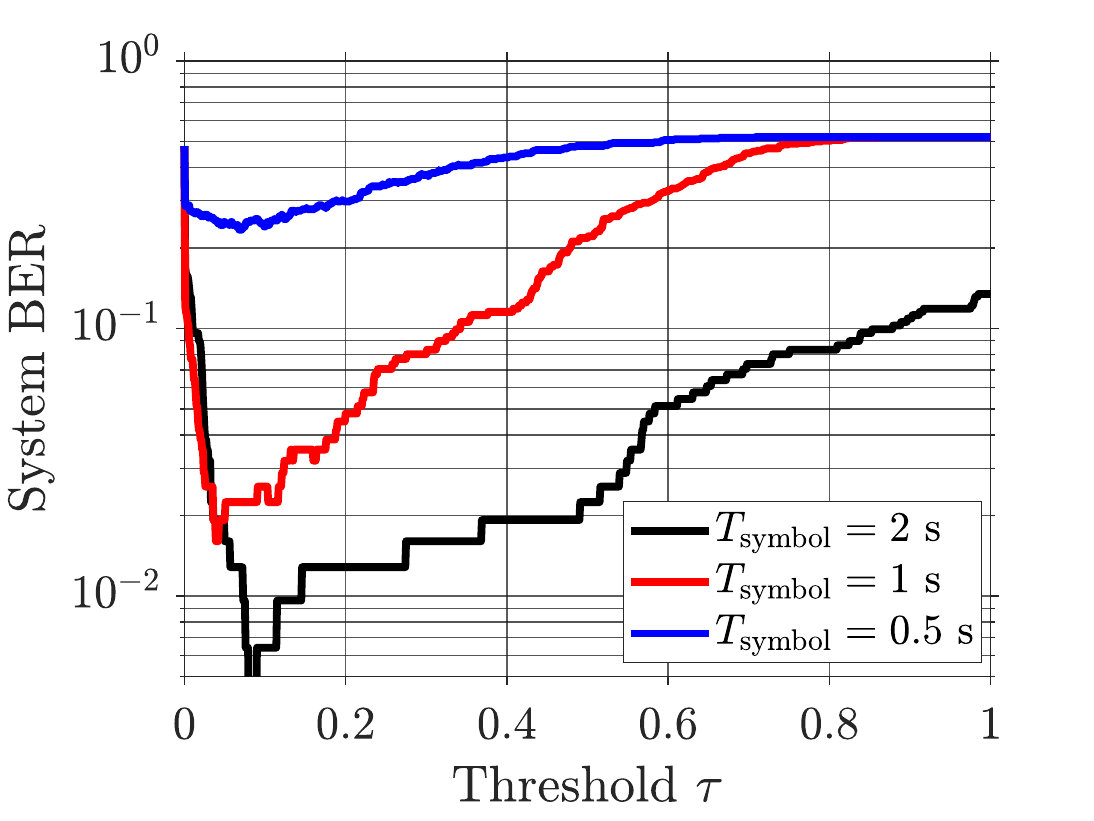}
    \caption{System \ac{BER} for a sample sequence of 312 bits sent via \ac{MUMO} transmission using a range of detection thresholds $\tau$, see Eq.~(\ref{eq:slope_detection}). Results for three different values of the symbol period $T_\mathrm{sym} = \{0.5, 1, 2\} \unit{\second}$ are depicted, representing a bit rate of 6, 3, and 1.5 \ac{bps}, respectively.}
    \label{fig:system_ber}
\end{figure}

The choice of threshold $\tau$ is fundamental in determining the resulting \ac{BER}. Therefore, Figure~\ref{fig:system_ber} depicts the resulting system \ac{BER} for $\tau \in [0,1]$. The evaluation was conducted for three different parameter settings of $\{T_\mathrm{sym}, T_\mathrm{inj}\}$, specifically $\{\qty{2}{\second}, \qty{0.1}{\second}\}$, $\{\qty{1}{\second}, \qty{0.05}{\second}\}$, and $\{\qty{0.5}{\second}, \qty{0.05}{\second}\}$ with a resulting bit rate of 1.5 \ac{bps}, 3 \ac{bps}, and 6 \ac{bps}, respectively. 
The results show almost error-free communication for $T_\mathrm{sym} = \qty{2}{\second}$ and $\qty{1}{\second}$ for the optimal choice of $\tau$, with 1 error and 5 errors across all 312 bits, respectively. In both cases, \ac{ISI} already plays a significant role, as we have seen in Figure~\ref{fig:avg_cir}, that the \ac{CIR} takes~$>\qty{2}{\second}$ to drop back to zero.
For $T_\mathrm{sym} = \qty{0.5}{\second}$, the \ac{BER} rises to about 23\% as \ac{ISI} and variations in the \ac{CIR} are difficult to overcome for the naive difference detector.

\begin{figure}[t]
    \centering
    \includeinkscape[width=\linewidth]{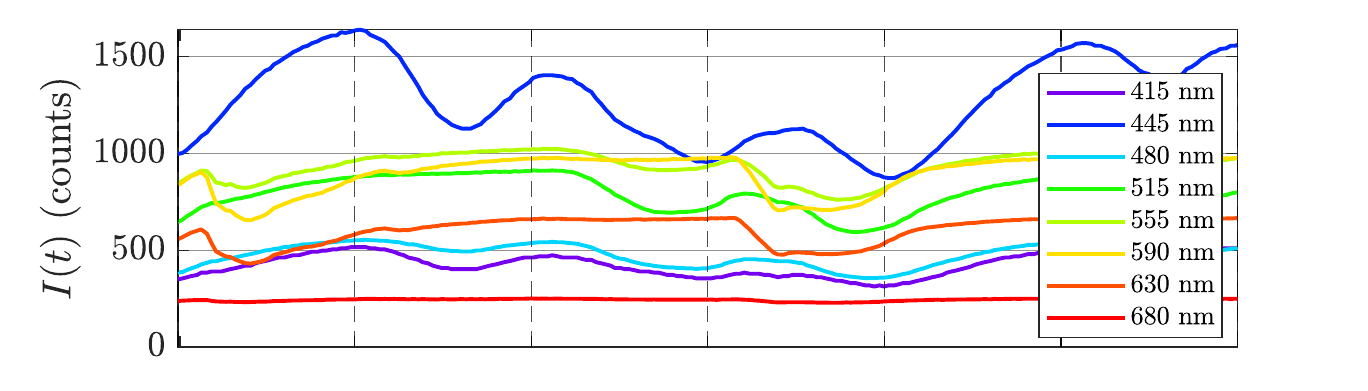}\\
    \includeinkscape[width=\linewidth]{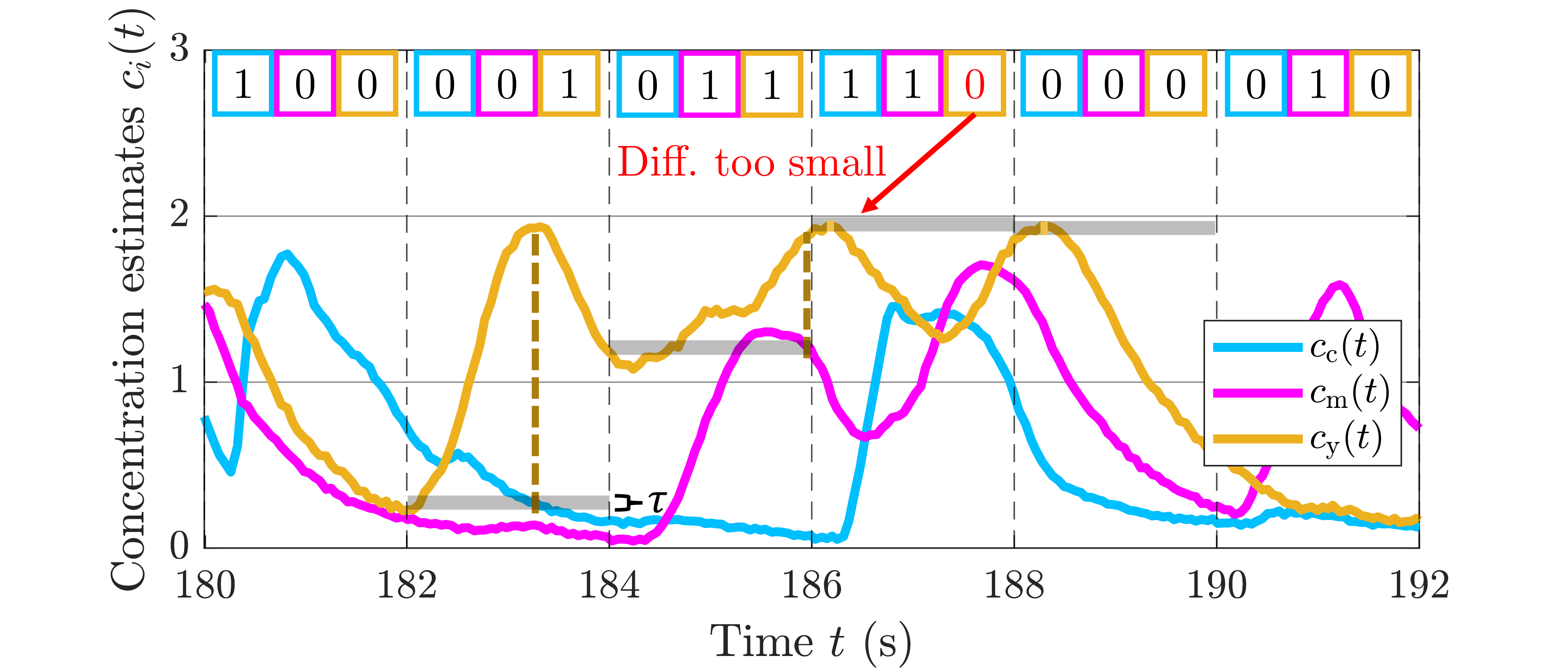}
    \caption{Selected time frame from the received signal during the random bit sequence transmission with $T_\mathrm{sym} = \qty{2}{\second}$. The plot on top depicts raw measured light intensities $I^j(t)$ for all channels, and the plot below depicts the corresponding estimated color intensities $c_i(t)$. Received bits in each symbol period are shown for the three inks at the top of the lower plot, with the single error of the entire sequence (yellow bit-1 transmitted, bit-0 detected) highlighted in red. For 4 selected symbol periods, the threshold $\tau$ and detected difference for the yellow signal are shown.}
    \label{fig:sequence}
\end{figure}

Figure~\ref{fig:sequence} shows an excerpt of the raw light intensity $I^j(t)$ in each channel and the estimation $c_i(t)$ during a random bit sequence transmission with $T_\mathrm{sym} = \qty{2}{\second}$. This side-by-side comparison showcases how the color intensity estimator transforms the measured $I(t)$ into the estimate by utilizing the (anti-)correlations between the different ink colors. Additionally, we can see the significant impact of \ac{ISI} on the shape of the pulses and that the difference detector can deal with the vast majority of distortions. The singular error of the 312-bit sequence, caused by particularly strong \ac{ISI} during a symbol with transmissions from all three colors, is highlighted in red.


\section{Conclusion}

This work introduced a novel flow-based \ac{MC} testbed design using spectral sensing and estimation procedures to differentiate between multiple color molecules, enabling real-time \ac{MUMO} transmission and reception. We presented inexpensive and easily operable \ac{TX} and information molecules and a non-invasive \ac{RX}. Due to custom 3D-printed injection and detection casings, easy extension towards \ac{MC} networks will be enabled. \Ac{CIR} results were presented, showing limited effects of the \ac{MUMO} on the signal. The preliminary performance evaluation indicates the potential of our real-time \ac{MUMO} platform to achieve single-digit \ac{bps} throughput even with a straightforward detector design. Related work has shown significant throughput gains using more sophisticated schemes~\cite{farsadNovelExperimentalPlatform2017, bartunikDevelopmentBiocompatibleTestbed2023}, and we plan to combine this with the demonstrated \ac{MUMO} gains in future work, as well as incorporate the broad possibilities of coding, modulation, and resource management schemes using multiple molecule types to improve communication performance further. An evaluation of multiple \acp{TX} and \acp{RX} is a crucial next step, and we plan to implement schemes, such as non-orthogonal multiple access~\cite{wietfeldDBMCNOMAEvaluatingNOMA2024} on the testbed. The real-time capabilities of the platform also enable feedback and control loops that could be crucial for evaluating various health-related sensing and actuation use cases.

\bibliographystyle{ieeetr}
\bibliography{references}

\begin{thebibliography}{10}

\bibitem{lotterExperimentalResearchSynthetic2023}
S.~Lotter, L.~Brand, V.~Jamali, M.~Schäfer, H.~M. Loos, H.~Unterweger, {\em et~al.}, ``Experimental {Research} in {Synthetic} {Molecular} {Communications} –{Part} {II},'' {\em IEEE Nanotechnol. Mag.}, vol.~17, no.~3, 2023.

\bibitem{farsadTabletopMolecularCommunication2013}
N.~Farsad, W.~Guo, and A.~W. Eckford, ``Tabletop {{Molecular Communication}}: {{Text Messages}} through {{Chemical Signals}},'' {\em PLoS ONE}, vol.~8, Dec. 2013.

\bibitem{hofmannTestbedbasedReceiverOptimization2022}
P.~Hofmann, J.~T. G{\'o}mez, F.~Dressler, and F.~H. Fitzek, ``Testbed-based {{Receiver Optimization}} for {{SISO Molecular Communication Channels}},'' in {\em Proceedings of the 2022 {{BalkanCom}}}, Aug. 2022.

\bibitem{kooMolecularMIMOTheory2016}
B.-H. Koo, C.~Lee, H.~B. Yilmaz, N.~Farsad, A.~Eckford, and C.-B. Chae, ``Molecular {{MIMO}}: {{From Theory}} to {{Prototype}},'' {\em IEEE JSAC}, vol.~34, Mar. 2016.

\bibitem{hofmannAnalogNetworkCoding2023}
P.~Hofmann, J.~Cabrera~Guerrero, R.~Bassoli, and F.~Fitzek, ``Analog {{Network Coding}} in {{Molecular Communications}}: {{A Practical Implementation}},'' in {\em Proceedings of the 2023 {{IEEE}} {GLOBECOM}}, IEEE, Dec. 2023.

\bibitem{farsadNovelExperimentalPlatform2017}
N.~Farsad, D.~Pan, and A.~Goldsmith, ``A {{Novel Experimental Platform}} for {{In-Vessel Multi-Chemical Molecular Communications}},'' in {\em Proceedings of the 2017 {{IEEE}} {GLOBECOM}}, Dec. 2017.

\bibitem{brandClosedLoopMolecular2024}
L.~Brand, M.~Scherer, M.~Sch{\"a}fer, A.~Burkovski, H.~Sticht, K.~Castiglione, and R.~Schober, ``Closed {{Loop Molecular Communication Testbed}}: {{Setup}}, {{Interference Analysis}}, and {{Experimental Results}},'' in {\em Proceedings of the {{IEEE International Conference}} on {{Communications}} ({{ICC}})}, June 2024.

\bibitem{angerbauerSalinityBasedMolecularCommunication2023b}
S.~Angerbauer {\em et~al.}, ``Salinity-{{Based Molecular Communication}} in {{Microfluidic Channels}},'' {\em IEEE Trans. Mol. Biol. Multi-Scale Commun.}, vol.~9, June 2023.

\bibitem{wickeExperimentalSystemMolecular2022}
W.~Wicke, H.~Unterweger, J.~Kirchner, L.~Brand, A.~Ahmadzadeh, D.~Ahmed, {\em et~al.}, ``Experimental {{System}} for {{Molecular Communication}} in {{Pipe Flow With Magnetic Nanoparticles}},'' {\em IEEE Trans. Mol. Biol. Multi-Scale Commun.}, vol.~8, June 2022.

\bibitem{bartunikDevelopmentBiocompatibleTestbed2023}
M.~Bartunik, G.~Fischer, and J.~Kirchner, ``The {{Development}} of a {{Biocompatible Testbed}} for {{Molecular Communication With Magnetic Nanoparticles}},'' {\em IEEE Trans. Mol. Biol. Multi-Scale Commun.}, vol.~9, June 2023.

\bibitem{walterRealtimeSignalProcessing2023}
V.~Walter, D.~Bi, A.~{Salehi-Reyhani}, and Y.~Deng, ``Real-time signal processing via chemical reactions for a microfluidic molecular communication system,'' {\em Nat. Commun.}, vol.~14, Nov. 2023.

\bibitem{akyildizInternetBioNanoThings2015}
I.~F. Akyildiz, M.~Pierobon, S.~Balasubramaniam, and Y.~Koucheryavy, ``The {I}nternet of {{Bio-Nano}} {T}hings,'' {\em IEEE Commun. Mag.}, vol.~53, Mar. 2015.

\bibitem{wangPracticalScalableMolecular2023}
J.~Wang, S.~{\"O}{\u g}{\"u}t, H.~Al~Hassanieh, and B.~Krishnaswamy, ``Towards {{Practical}} and {{Scalable Molecular Networks}},'' in {\em Proceedings of the 2023 {{ACM SIGCOMM}}}, Sept. 2023.

\bibitem{panMolecularCommunicationPlatform2022}
W.~Pan, X.~Chen, X.~Yang, N.~Zhao, L.~Meng, and F.~H. Shah, ``A {{Molecular Communication Platform Based}} on {{Body Area Nanonetwork}},'' {\em Nanomaterials}, vol.~12, Feb. 2022.

\bibitem{caliExperimentalImplementationMolecule2024a}
F.~Cal{\`i}, S.~Barreca, G.~{Li-Destri}, A.~Torrisi, A.~Licciardello, and N.~Tuccitto, ``Experimental {{Implementation}} of {{Molecule Shift Keying}} for {{Enhanced Molecular Communication}},'' {\em IEEE Trans. Mol. Biol. Multi-Scale Commun.}, vol.~10, Mar. 2024.

\bibitem{bartunikColourspecificMicrofluidicDroplet2020}
M.~Bartunik, M.~Fleischer, W.~Haselmayr, and J.~Kirchner, ``Colour-specific microfluidic droplet detection for molecular communication,'' in {\em Proceedings of the 7th {{ACM}} {{NanoCom}}}, Oct. 2020.

\bibitem{wickeMagneticNanoparticleBasedMolecular2019}
W.~Wicke, A.~Ahmadzadeh, V.~Jamali, H.~Unterweger, C.~Alexiou, and R.~Schober, ``Magnetic {{Nanoparticle-Based Molecular Communication}} in {{Microfluidic Environments}},'' {\em IEEE Trans. NanoBiosci.}, vol.~18, Apr. 2019.

\bibitem{swinehartBeerLambertLaw1962}
D.~F. Swinehart, ``The {{Beer-Lambert Law}},'' {\em Journal of Chemical Education}, vol.~39, July 1962.

\bibitem{QuantitiveAnalysis2007}
P.~R. Griffiths and J.~A. De~Haseth, ``Quantitative {A}nalysis,'' in {\em Fourier {Transform} {Infrared} {Spectrometry}}, New York, U.S.: Wiley, 1.~ed., 2007.

\bibitem{kayFundamentalsStatisticalSignal1993}
S.~M. Kay, {\em Fundamentals of {{Statistical Signal Processing}}: {{Estimation Theory}}}.
\newblock New Jersey, U.S.: Prentice-Hall, Inc., 1993.

\bibitem{jamaliChannelModelingDiffusive2019}
V.~Jamali, A.~Ahmadzadeh, W.~Wicke, A.~Noel, and R.~Schober, ``Channel {{Modeling}} for {{Diffusive Molecular Communication}}\textemdash{{A Tutorial Review}},'' {\em Proc. IEEE}, vol.~107, July 2019.

\bibitem{whiteFluidMechanics2016}
F.~M. White, {\em Fluid Mechanics}.
\newblock New York, U.S.: McGraw-Hill, 2016.

\bibitem{holzTemperaturedependentSelfdiffusionCoefficients2000}
M.~Holz, S.~R. Heil, and A.~Sacco, ``Temperature-dependent self-diffusion coefficients of water and six selected molecular liquids for calibration in accurate {{1H NMR PFG}} measurements,'' {\em Royal Society of Chemistry PCCP Journal}, vol.~2, no.~20, 2000.

\bibitem{wietfeldDBMCNOMAEvaluatingNOMA2024}
A.~Wietfeld, S.~Schmidt, and W.~Kellerer, ``{{DBMC-NOMA}}: {{Evaluating NOMA}} for {{Diffusion-Based Molecular Communication Networks}},'' in {\em Proceedings of the 2024 {{IEEE}} {{ICC}}}, June 2024.

\end{thebibliography}

\end{document}